\documentclass[showpacs,twocolumn,preprintnumbers,amsmath,amssymb,prl,epsf,superscriptaddress]{revtex4-1}

\usepackage{graphicx}
\usepackage{nicefrac}
\usepackage{color}
\usepackage{lineno}
\usepackage{setspace}

\newcommand{\R}{\color{black}}

\begin{document}
\title{Overcoming detection loss and noise in squeezing-based optical sensing}
\author{Gaetano Frascella}
\affiliation{Max Planck Institute for the Science of Light, Staudtstra\ss{}e 2, 91058 Erlangen, Germany}
\affiliation{University of Erlangen-N\"urnberg, Staudtstra\ss{}e 7/B2, 91058 Erlangen, Germany}
\author{Sascha Agne}
\affiliation{Max Planck Institute for the Science of Light, Staudtstra\ss{}e 2, 91058 Erlangen, Germany}
\author{Farid Khalili}
\affiliation{Russian Quantum Center, Bolshoy Bulvar 30/bld. 1, 121205 Skolkovo,
Moscow, Russia.}
\affiliation{NUST ``MISiS'', Leninskiy Prospekt 4, 119049 Moscow, Russia.}
\author{Maria~V.~Chekhova}
\email{maria.chekhova@mpl.mpg.de}
\affiliation{Max Planck Institute for the Science of Light, Staudtstra\ss{}e 2, 91058 Erlangen, Germany}
\affiliation{University of Erlangen-N\"urnberg, Staudtstra\ss{}e 7/B2, 91058 Erlangen, Germany}

\maketitle \narrowtext

\begin{spacing}{1}

\textbf{Among the known resources of quantum metrology, one of the most practical and efficient is squeezing. Squeezed states of atoms and light improve the sensing of the phase, magnetic field, polarization, mechanical displacement. They promise to considerably increase signal-to-noise ratio in imaging and spectroscopy, and are already used in real-life gravitational-wave detectors. But despite being more robust than other states, they are still very fragile, which narrows the scope of their application. In particular, squeezed states are useless in measurements where the detection is inefficient or the noise is high. Here, we experimentally demonstrate a remedy against loss and noise: strong noiseless amplification before detection. This way, we achieve loss-tolerant operation of an interferometer fed with squeezed and coherent light. With only 50\% detection efficiency and with noise exceeding the level of squeezed light more than 50 times, we overcome the shot-noise limit by 6~dB. Sub-shot-noise phase sensitivity survives up to 87\% loss. Application of this technique to other types of optical sensing and imaging promises a full use of quantum resources in these fields.}

Quantum resources promise advances in sensing and metrology~\citep{Giovannetti:04,Mitchell:04,Nagata:07,Resch:07,Degen2017,Pezze2018,Wolf:19}, beyond the fundamental limits of precision set for classical light. They enable overcoming the shot-noise limit (SNL) and achieving the ultimate Heisenberg limit in phase sensing~\citep{Demkowicz:15}. Within the quantum states toolbox, most practical are squeezed states. Unlike exotic non-Gaussian states, squeezed states of atoms and light can contain a macroscopic number of particles and survive a reasonable amount of loss while providing reduced uncertainty in a plethora of measurements. As a result, their use {\color{black} noticeably improved the sensitivity of modern gravitational-wave detectors}~\citep{LIGO:11,Barsotti:18,Tse:19,Acernese_PRL_123_231108_2019}. There are also numerous proof-of-principle experiments on squeezing-enhanced  absorption measurement and spectroscopy~\citep{Polzik1992,Whittaker2017,Moreau2017}, imaging and microscopy~\citep{Brida2010,Taylor:14,Samantaray2017,SabinesChesterking2019}, polarimetry~\citep{Feng2004}, magnetometry~\citep{Wolfgramm2010,Lucivero2016}, and other types of sensing~\citep{Treps:02,Lawrie:19}.

Although more robust than non-Gaussian states, squeezed states are still very susceptible to loss. Therefore, overcoming the SNL in optical sensing requires extremely efficient detection~\citep{Eberle2010,Samantaray2017}. The quantum advantage provided by squeezing strongly depends on the detection efficiency $\eta$~\citep{Demkowicz:13,Manceau:17,Knyazev2019}, and completely disappears for $\eta<50\%$. Similarly, detection noise becomes a problem if the probing state is not very bright. In spectral ranges where detection is inefficient or noise is high, sub-shot-noise optical sensing is impossible.

Here we experimentally demonstrate a solution to this problem, namely strong parametric amplification of the signal before detection. Although proposed long ago~\citep{Caves:81} and already applied to the detection of microwave quantum states~\citep{Mallet2011},  {\R to the improvement of the homodyne detection bandwidth~\citep{Shaked:18}}, to the tomography of optical non-Gaussian states~\citep{LeJeannic2018}, and to atom interferometry~\citep{Hosten2016}, this method has not been implemented in optical sensing. A similar principle provides loss tolerance of so-called SU(1,1) (nonlinear) interferometers~\citep{Yurke:86,Ou:12,Manceau:17PRL}, both optical and atomic. But SU(1,1) has certain limitations, among them, a narrow range of phase sensitivity~\citep{Manceau:17} and more complexity in realization, \textcolor{black}{at least in the multimode version}. For making phase measurement loss- and noise-tolerant, much more practical is to add an amplifier to the output of a  linear [SU(2)] interferometer. This is what we do in our experiment; the same method can be applied to sub-shot-noise imaging and absorption measurement.

\section{Results}

\begin{figure}[h]
\includegraphics[width=0.96\columnwidth]{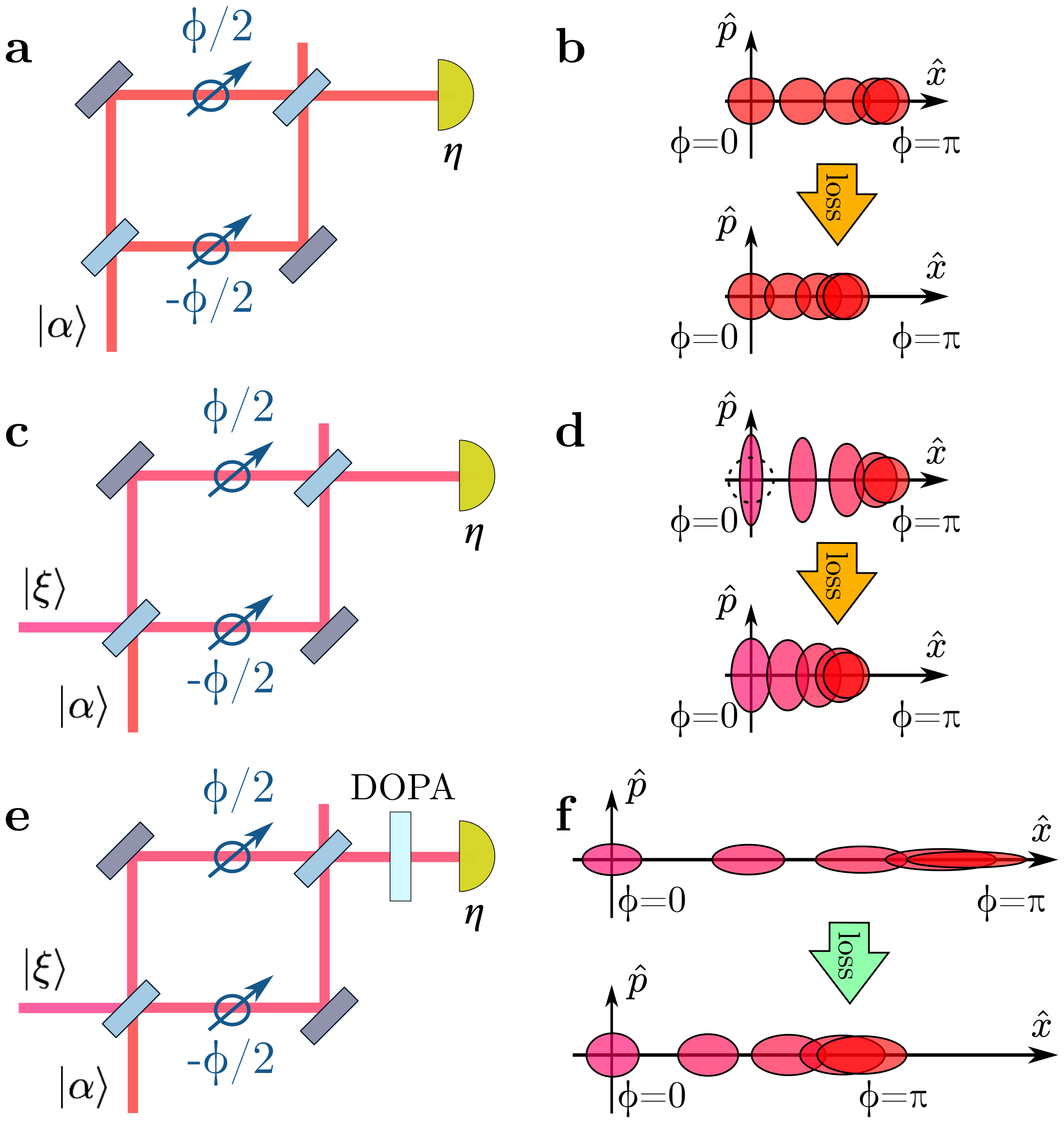}\caption{\label{fig:su2withsv} \textbf{Overcoming detection loss in squeezing-assisted interferometry.} Panels a,c,e show the experimental schemes and panels b,d,f, the calculated Wigner functions of the output states for five equidistant phases. \textbf{a:} {\R This ideal} Mach-Zehnder interferometer fed with a coherent state $\left|\alpha\right>$ has the best phase sensitivity at $\phi=0$, the detector seeing the `dark fringe', where the output states for different phases are most distinguishable (\textbf{b}). \textbf{c:} If, additionally, the second input port is fed with squeezed vacuum (SV) $\left|\xi\right>$, the neighbouring states are squeezed and distinguishable better; however loss or imperfect detection efficiency $\eta$ makes them overlap (\textbf{d}).  \textbf{e:} To overcome the loss, a degenerate optical parametric amplifier (DOPA) amplifies the state before detection. \textbf{f:} The output states are now anti-squeezed in the quadrature carrying the phase information; the latter is therefore protected against loss. In the calculation, $\alpha=3,\, \eta=0.5$, $\left|\xi\right>$ is $6$~dB squeezed, and DOPA provides $9.6$~dB quadrature anti-squeezing.}
\end{figure}

The principle is illustrated in Fig.~\ref{fig:su2withsv} showing a Mach-Zehnder interferometer as an example. In classical interferometry (a), coherent light is fed into one input port, and direct or homodyne detection is performed in one output port in order to sense a phase difference $\phi$ between the two arms. The best sensitivity is achieved in the `dark fringe', with no light in the output port. This becomes clear by looking at the Wigner function of the output state. In panel b, it is plotted for
$\phi={\R 0,\, 0.2\pi,\textrm{.. }\pi}$
for the input coherent state containing $9$ photons. Whenever the output states are well separated, the phase sensitivity is high; this happens around $\phi=0$. The phase can be retrieved by measuring the $\hat x$ quadrature (homodyne detection) or the number of photons (direct detection).

To increase the phase sensitivity, a squeezed vacuum state $\left|\xi\right>$ is injected into the unused input port~\citep{Caves:81,Xiao:87,Grangier1987,Schafermeier:18} (Fig.~\ref{fig:su2withsv} c). Then, under the `dark fringe' condition $\phi=0$, the detected output port contains the squeezed vacuum ($6$~dB squeezing is assumed). As the phase changes from $\phi=0$ to $\phi=\pi$, the output state evolves to a coherent state (Fig.~\ref{fig:su2withsv} d). Near the `dark fringe', the states are squeezed in the $\hat x$ quadrature and therefore better distinguishable. However, if the detection is lossy (we assume $\eta=0.5$), the squeezing is degraded, the states overlap, and the advantage of squeezing is fully or partly lost.

As a remedy, the state at the output of the interferometer is amplified by a phase sensitive amplifier (a degenerate optical parametric amplifier, DOPA), see Fig.~\ref{fig:su2withsv} e. The DOPA amplifies the quadrature carrying the phase information, but it also makes this quadrature anti-squeezed. The relative separation of the states at the output (Fig.~\ref{fig:su2withsv} f) remains the same, but the phase information is now in the anti-squeezed quadrature and therefore less susceptible to loss. In the calculation, quadrature anti-squeezing of $9.6$~dB completely eliminates the effect of loss; as we will see below, experiment provides even stronger amplification and a much higher loss can be overcome. If strong enough, amplification also protects the measurement against the detection noise. Importantly, a DOPA does not change the signal-to-noise ratio of the input state and therefore does not add noise.

In our experiment, we provide strong phase-sensitive parametric amplification with a nonlinear beta-barium borate (BBO) crystal pumped by picosecond pulses under collinear frequency-degenerate type-I phasematching. The amplification is defined by the squeeze factor $G\propto\chi^{(2)}L\sqrt{P}$, where $L$ is the length of the nonlinear crystal, $\chi^{(2)}$ its second-order susceptibility, and $P$ the pump power~\citep{Iskhakov:12}. The quadratures after such an amplifier evolve as {\R $\hat x_{\rm out}=e^{G} \hat x_{\rm in},\, \hat p_{\rm out}=e^{-G} \hat p_{\rm in}$}, and the mean photon number of a coherent state, as  $N_{\rm out}=N_{\rm in}e^{2G}+\sinh^2G$. Generally, we can achieve up to $G\sim10$~\citep{Iskhakov:12}. The current experiment uses squeeze factors only up to {\R $G=3.6$}, which still means impressive {\R $1340$}-fold intensity amplification and  {\R $31$}~dB quadrature variance anti-squeezing.

\begin{figure}[b]
\includegraphics[width=\columnwidth]{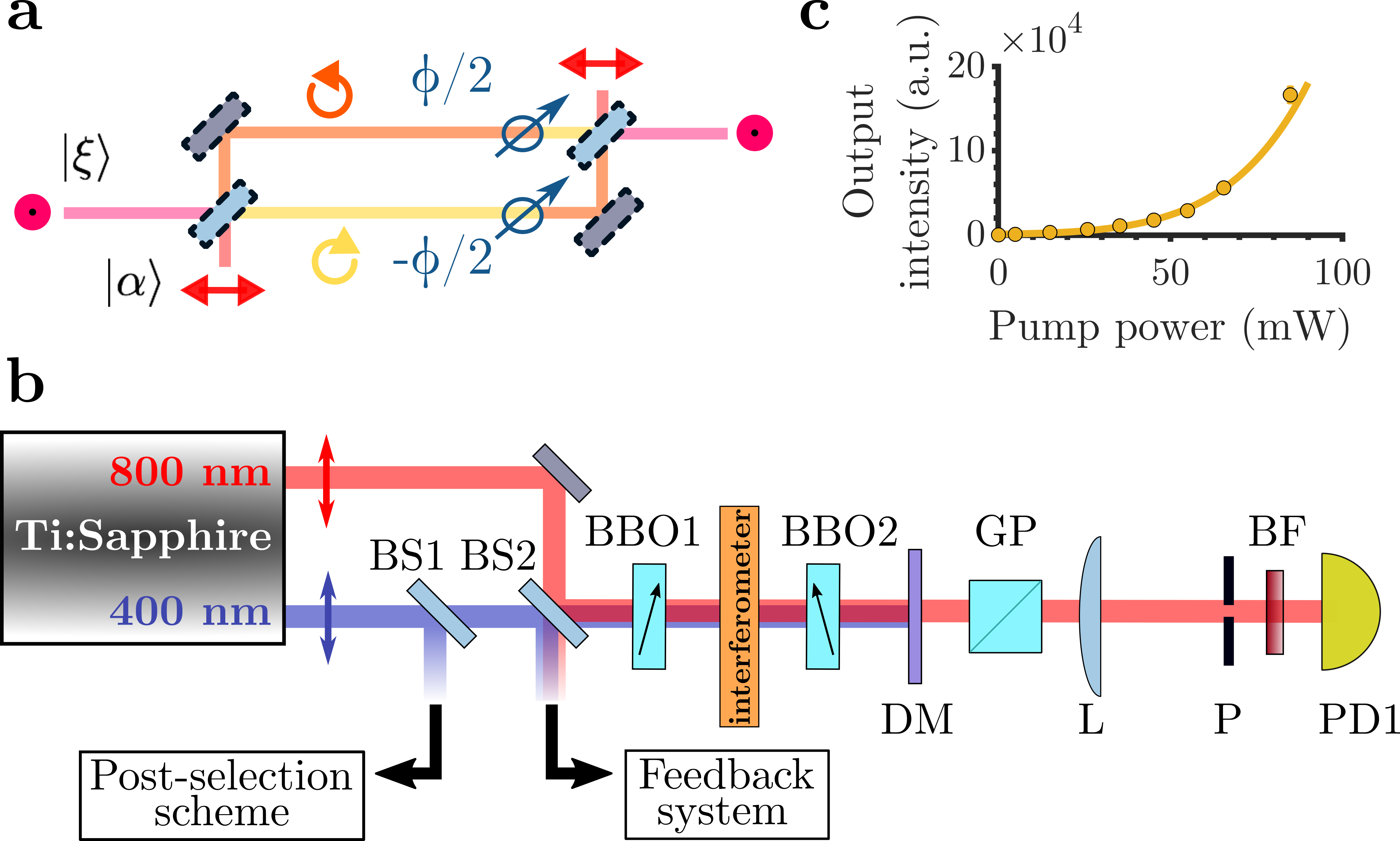}\caption{\label{fig:setup} \textbf{Squeezing-assisted interferometer with parametric amplification at the output.} \textbf{a:} A HWP as a Mach-Zehnder interferometer: left- and right-circularly polarized states play the role of the two arms and linear polarization states, input and output modes. \textbf{b:} Vertically polarized {\R squeezed vacuum (SV)} is produced in nonlinear crystal BBO1 and the horizontally polarized coherent state is fed through beamsplitter BS2. Vertically polarized output of the interferometer is amplified in the second crystal BBO2, cleaned from the pump by dichroic mirror DM, selected by Glan polarizer GP, and finally detected with photodetector PD1. Lens L and pinhole P provide spatial filtering, while bandpass filter BF, spectral filtering. Beamsplitters BS1, BS2 tap off parts of the pump and coherent beams to monitor intensity fluctuations and lock the phase. \textbf{c:} The output intensity of unseeded BBO2 versus the pump power.}
\end{figure}
In our proof-of-principle implementation, the interferometer is \textcolor{black}{simulated by} a half-wave plate (HWP)~\citep{Schafermeier:18}, see Fig.~\ref{fig:setup} a. Instead of two spatial modes, the two arms of the interferometer are the right- and left-circularly polarized modes, and rotation of the HWP leads to a phase shift between them (see Methods). The inputs and outputs of the interferometer are linear polarization modes. The coherent state is the radiation of a Spectra Physics Spitfire Ace laser with the central wavelength $800$ nm, $1.5$ ps pulse duration, $5$ kHz repetition rate, horizontally polarized (Fig.~\ref{fig:setup} b), attenuated to $1500$ photons/pulse. The {\R squeezed vacuum (SV)}, produced in the first amplifier BBO1 pumped by the second harmonic of the same laser and unseeded, is vertically polarized. The coherent beam, injected through beamsplitter BS2, does not interact with BBO1 because {\R there is no phase matching for its} polarization. The output amplifier BBO2 amplifies only the vertical polarization mode~\footnote{The {\R de-amplification} phase is set by properly choosing the distance between the two crystals~\cite{Manceau:17PRL}{\R, while the amplification phase of the coherent beam is set with a piezoelectric actuator (Sec. II of the Supplemental Material).}}. The pump is rejected with dichroic mirror DM and the vertical polarization is selected by Glan polarizer GP. To ensure {\R good de-amplification of the SV}, we select an angular bandwidth of $130{\rm \,\mu rad}$ with $200{\rm-\mu m}$ pinhole P in the focal plane of lens L ($f=1.5{\rm \,m}$). The bandpass filter BF has $3$ nm spectral bandwidth around the central wavelength $800$ nm. Photodetector PD1 registers the number of photons per pulse, with a dark noise of $500$ photons.

The loss between BBO1 and BBO2 (internal) amounts to $3\%$ and is impossible to compensate for. The loss after the amplifier (`detection loss') is incorporated into the detection efficiency $\eta=0.50\pm0.03$ (for the detailed description of losses, see Sec. III of the Supplemental Material). {\R Please note that spatial and spectral filtering, i.e. the pinhole and the bandpass filter, restricts the amount of photons detected but cannot be considered as loss since all the measurements are carried out within the same bandwidths. Meanwhile, the optical transmission of the bandpass filter within the spectral bandwidth is taken into account.} The amplification at the interferometer output, if strong enough, can completely overcome {\R the detection inefficiencies}. Both the amplification of BBO2 and the squeezing of BBO1 are characterized by measuring the photon numbers at their outputs  with a vacuum at the input (see Methods and Fig.~\ref{fig:setup} c). In all measurements, $G_1=1.7\pm0.3$  {\R(measured as described in Methods)}, {\R from which an initial $15$~dB level of squeezing is inferred}.

Beamsplitters BS1 and BS2 serve, respectively, for monitoring/stabilizing the pump intensity (Methods) and locking the phase between the coherent and SV beams at the input of the HWP (Sec. II of the Supplemental Material).

The phase sensitivity is evaluated as
\begin{equation}
\Delta\phi=\Delta N\biggl(\left|\frac{d\left<N\right>}{d\phi}\right|\biggr)^{-1},\label{eq:phasesens}
\end{equation}
where $\Delta N=\sqrt{\left<N^{2}\right>-\left<N\right>^{2}}$ is the measured uncertainty of the photon number, detector dark noise included, and $\left<N\right>$ is the average number of photons. The slope of the $\left<N\right>\left(\phi\right)$ dependence is inferred from the fit.
\begin{figure}
\includegraphics[width=.9\columnwidth]{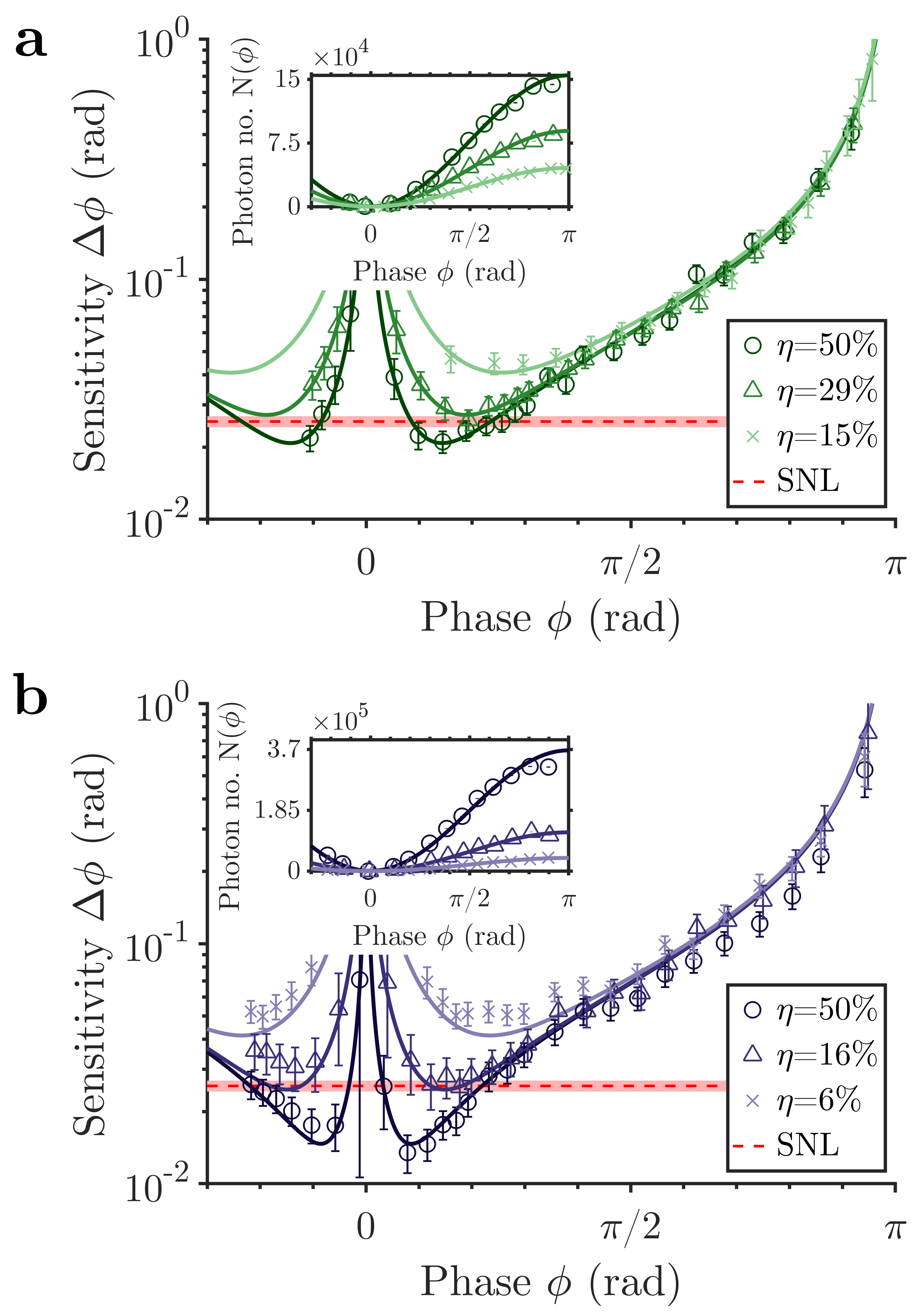}\caption{\label{fig:phase-sens}\textbf{Phase sensitivity measurements with $15$~dB input squeezing and $N_{\alpha}=1500$ photons in the coherent beam.} The red dashed line marks the SNL and the solid lines are the theoretical predictions. The insets show the photon number dependence on the phase. \textbf{a:}  Parametric amplification with $G_2=2.7$. Circles, {\R triangles and crosses} show measurements with detection efficiency $\eta=50\%,\,29\%$ and $15\%$, respectively.
\textbf{b:} Parametric amplification with $G_2=3.2$. Circles, {\R triangles and crosses} are for $\eta=50\%,\,16\%$ and $6\%$, respectively. }
\end{figure}

Figure~\ref{fig:phase-sens} shows the results of phase sensitivity measurements, with the SNL (measured as described in Methods) marked with a dashed red line. In panel a, the second amplifier has \textcolor{black}{the squeeze factor measured to be $G_2=3.1\pm0.3$ (photon-number amplification $490$ times)} and the detection efficiency is $50\%,\,29\%$ and $15\%$. To vary $\eta$, we place a HWP and a GP (not shown in Fig.~\ref{fig:setup}) before lens L. For the highest efficiency (circles), the SNL variance is overcome by $1.8\pm0.8$~dB in the best case, i.e. at $\phi\sim0$. Note that without the output amplifier, the phase sensitivity would not overcome the SNL at \textcolor{black}{all, in particular due to the large detection noise}. With the amplifier, the performance just overcomes the SNL for $29\%$ detection efficiency ({\R triangles}). Exactly at $\phi=0$, the average number of photons is the lowest and the detector dark noise spoils the sensitivity, hence the peak.

By increasing the output parametric amplification (\textcolor{black}{the measured squeeze factor $G_2=3.6\pm0.3$}), the phase sensitivity is improved (Fig.~\ref{fig:phase-sens} b). The measurements shown with circles, {\R triangles and crosses} correspond, respectively, to detection efficiencies $\eta=50\%,\,16\%$ and $6\%$. For the highest value of $\eta$, the SNL variance is overcome by $6\pm1$~dB.
In addition, with {\R lower detection efficiency} we need a {\R stronger amplification} for overcoming SNL. Indeed, while it is $\eta=29\%$ for lower $G_2$ (a), for higher $G_2$ it is only $\eta=16\%$ (b).

To fit the experimental points, we derive the phase sensitivity for a lossy squeezing-assisted interferometer with an OPA at the output and direct detection~\citep{Manceau:17}. The model also takes into account the relative excess noise of the coherent beam, i.e. deviation of the normalised second-order correlation function $g^{\left(2\right)}$ from unity (see Secs. IV-V of the Supplemental Material){\R, which we measured to be $0.0020\pm0.0005$}. We use two fitting parameters $G_{2}$ and $g^{\left(2\right)}$ for the model (lines), in \textcolor{black}{reasonable} agreement with the experiment for $G_{2}=2.7$ and $g^{\left(2\right)}=1.003$ (a) and for $G_{2}=3.2$ and $g^{\left(2\right)}=1.004$ (b).

Ideally, the sub-shot-noise sensitivity phase range should cover half of the $2\pi$ period~\citep{Manceau:17}. In our case, it is at most $\sim0.3\,\pi$ in panel a and $\sim0.4\,\pi$ in panel b. The main reason is that our coherent beam is not shot-noise limited. Additionally, the detector dark noise excludes part of this range near $\phi=0$. But even with these imperfections, the sub-shot-noise sensitivity range is broader than the one of the SU(1,1) interferometer~\citep{Manceau:17PRL,Anderson:17}.

\begin{figure}
\includegraphics[width=.9\columnwidth]{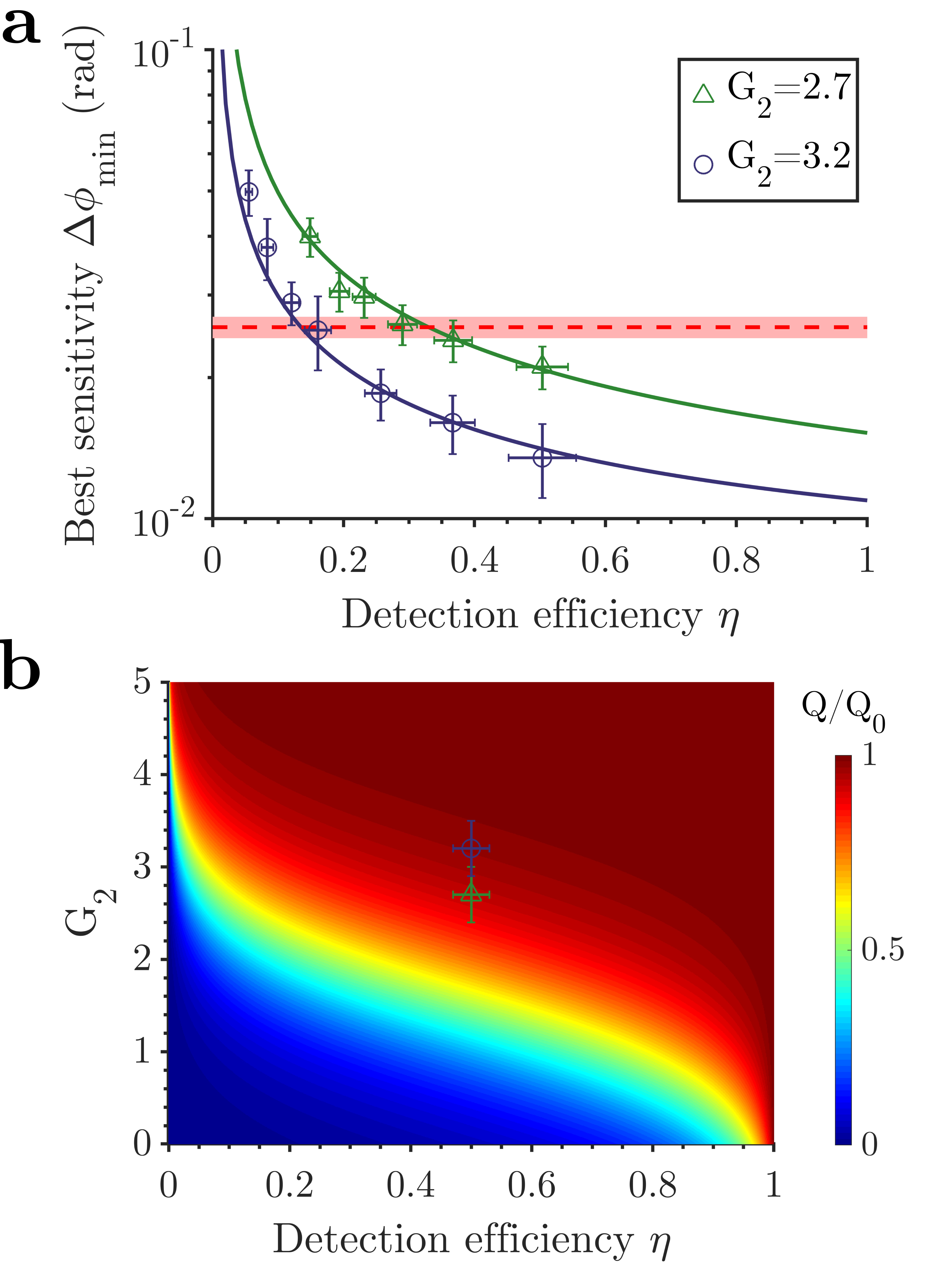}\caption{\label{fig:loss}\textbf{Loss tolerance through phase-sensitive amplification. a} Best phase sensitivity as a function of detection efficiency $\eta$ with $G_2=3.2$ and $G_2=2.7$ amplification (blue circles and green {\R triangles}, respectively). The SNL is the red dashed line and the lines show the theoretical model. \textbf{b} The quantum advantage $Q$ normalized to the 'perfect detection' one $Q_0=17$ versus the detection efficiency and parametric amplification. Points show the experimental parameters used.}
\end{figure}

To demonstrate tolerance to detection inefficiency, Fig.~\ref{fig:loss} a shows the best phase sensitivity $\Delta\phi_{{\rm min}}$ versus the detection efficiency $\eta$. The photon number for the coherent beam and the input squeezing are kept constant; the common SNL is marked with a red dashed line. We observe that with $G_{2}=2.7$ (green {\R triangles}), the system provides sub-shot-noise sensitivity for $\eta>34\%$. For $G_{2}=3.2$ (blue circles), the best sensitivity proves more robust and overcomes the SNL for $\eta=13\%$.

For any detection efficiency, by increasing $G_2$ one can get as close as necessary to perfect-detection performance. Indeed, with an account for amplification, the quantum advantage in phase sensitivity is
\begin{equation}
  Q = \biggl(Q_0^{-1} + \frac{1-\eta}{\eta\mu}e^{-2G_2}\biggr)^{-1},
\end{equation}
where $Q_0$ is the quantum advantage under perfect detection~\citep{Manceau:17} {\R and $(1-\mu)$ the internal loss}. Figure~\ref{fig:loss} b shows $Q/Q_0$, where $Q_0=17$ is determined by the initial $15$~dB squeezing and $3$\% internal loss, as a function of $\eta$ and $G_2$. Blue and green points mark the parameters of our experiment. For very low detection efficiency, the required amplification can be unrealistically high, but $G=5$, easily achievable in experiment~\citep{Iskhakov:12,Manceau:17PRL}, already enables overcoming $98\%$ loss.

Noteworthy, the detection of pulsed light we perform here {\R has a considerable noise, equivalent to $500$ photons per pulse. Yet, although this noise is much stronger than the SV (about $10$ photons per pulse) and comparable to the coherent beam ($1500$ photons per pulse) injected into the interferometer, parametric amplification  provides noise robustness and} enables sub-shot-noise phase sensitivity. Meanwhile, it is the detection noise that prevents reaching the `perfect detection' quantum advantage (green and blue points in Fig.~\ref{fig:loss} b).

\section{Discussion}
We have demonstrated that for a squeezing-assisted interferometer the detrimental effects of detection loss and noise can be eliminated with a phase-sensitive parametric amplifier at the output. In particular, our experiment shows phase sensitivity overcoming the SNL by $6$~dB even with $50\%$ detection efficiency and noise comparable to the signal sensing the phase. Increasing the output amplification helps to overcome a higher amount of loss. By amplifying the output photon number about $600$ times we overcome the SNL for detection efficiency down to $13\%$. Even lower detection efficiencies will not be an obstacle for squeezing-assisted measurements if stronger amplification is used; it is realistic to go down to $2$\% efficiency.

This result is relevant to many schemes where losses in the output optical path, including limited detection efficiency, reduce the advantage brought by squeezing. {\color{black} An example is the gravitational-wave detectors, where, due to the output losses, 7-10~dB of the input squeezing gives only $\approx3$~dB of the sensitivity gain}~\citep{Tse:19,Acernese_PRL_123_231108_2019}. {\color{black} Compensating for these losses with a continuous-wave scheme similar to ours could  revolutionize gravitational-wave detection (note that the 6~dB gain in the phase sensitivity translates to almost an order of magnitude increase of the detection rate)}. Importantly, the homodyne scheme (used in gravitational-wave detectors) will also benefit from pre-amplification; moreover, unlike direct detection, it is not affected by detector dark noise~\citep{Manceau:17}.

A similar amplification strategy will be valid for imaging experiments involving multimode radiation, considered theoretically in~\citep{Knyazev2019}. Indeed, the scheme could be accommodated to support more than a single spatial mode, similarly to Ref.~\citep{Frascella:19}, and this opens up a considerable amount of experiments on sub-shot-noise imaging. The method will be also applicable to sub-shot-noise measurement of very small loss, where otherwise a very high detection efficiency is required~\citep{Losero2018,Knyazev2019}.

Detection efficiency proves to be an especially important constraint for experiments with mid-infrared and terahertz radiation, used for biological and industrial applications. Another case is where the whole angular spectrum of radiation cannot be collected due to small detector sizes. This is an issue in sub-shot-noise microscopy where high resolution compromises the detection efficiency~\citep{Samantaray2017}. In these cases, parametric amplification before detection will be indispensable.

\setlength\linenumbersep{.2cm}
\linenumbers
\section{Methods}
{\bf Polarization Mach-Zehnder interferometer} is implemented using a single HWP (Fig.~\ref{fig:setup} a). The input coherent state $\left|\alpha\right>$ has horizontal polarization and the SV state $\left|\xi\right>$, vertical polarization. The 50:50 beam splitters correspond to the transition between linear and circular polarization bases. The HWP with the optic axis at an angle $\delta$ introduces a phase shift $\phi=4\delta$ between the two arms, see Sec.~V of the Supplemental Material. The HWP is a 45-${\rm \mu m}$-thick dual-wavelength waveplate: a HWP for $800$ nm; a full-wave plate for $400$ nm.

{\bf Parametric amplifiers} BBO1 and BBO2 are 2 mm BBO crystals pumped by the second harmonic of Spectra Physics Spitfire Ace laser (central wavelength $400$ nm, horizontal polarization, average power $65$ mW, waist intensity FWHM $240\pm10{\rm \,\mu m}$). Their squeeze factors $G_{1,2}$ are found by measuring the mean numbers of photons $N_{1,2}$ per pulse at their outputs in the regime of vacuum amplification (high-gain parametric down-conversion) and fitting {\R with the parameters $B_{1,2}$} the obtained dependences on the pump power $P$ with the functions $N_{1,2}\propto\sinh^2 G_{1,2}$, $G_{1,2}=B_{1,2}\sqrt{P}$. By detuning BBO1 from exact phase matching, $G_1$ is reduced compared to $G_2$. {\R The waist of the coherent beam is inside BBO2 and the waist intensity FWHM is chosen close to the one of the first Schmidt mode~\citep{Wasilewski:06}, i.e. $80\pm10{\rm \,\mu m}$. This ensures mode matching for the coherent beam and, as a result, efficient amplification. Meanwhile, the SV is de-amplified efficiently only in the collinear direction due to its divergence, and spatial filtering at the detection stage is necessary for this purpose.}

 For {\bf pump intensity stabilization,} we tap off part of the pump beam by BS1 (Fig.~\ref{fig:setup} b), amplify the fluctuations (initially $2\%$ RMS) approximately 7 times using high-gain parametric down-conversion, and post-select measurements for which the pump fluctuations were within $0.3\%$ RMS (see Sec. I of the Supplemental Material).

{\bf The shot-noise limit for the phase sensitivity} is found as
\begin{equation}
\Delta\phi_{{\rm SNL}}=\frac{1}{\sqrt{N_{\alpha}+N_{\rm{SV}}}},
\end{equation}
where $N_{\alpha}$ and $N_{\rm{SV}}$ are, respectively, the photon numbers of the coherent and SV beams inside the interferometer, i.e. at the HWP, within the spatial and spectral bandwidths registered at the detector. For the SV beam, we can use the estimate $N_{\rm{SV}}\sim\sinh^2G_1$. For the coherent beam, we measure with the photo-detector at the output of the interferometer an average number of photons per pulse of $720\pm20$, whose uncertainty is reduced by repeated measurements. Dividing this number by the total efficiency $49\pm3\%$, we obtain $N_{\alpha}=1500\pm100$ and, neglecting the small value of $N_{\rm{SV}}$, the SNL sensitivity is $\Delta\phi_{{\rm SNL}}=26\pm1$ mrad.

We also test the interferometer in the classical regime, by blocking the SV and leaving only the laser beam at the input. In this case, the phase sensitivity is much worse than  SNL, due to the detection loss and noise (see Sec. III of the Supplemental Material). We note that it is this test that identifies the SNL in most of experiments. While it might work for continuous-wave lasers, it leads to the overestimation of the SNL for pulsed light, which usually has stronger intensity fluctuations.

{\R {\bf The error bars} reported in this work represent the standard error of the mean with a sample size of 1000. For the measurement of the number of photons, the error bars are too small to be seen.}

\section{Supplementary information}
Supplementary information to ``Overcoming detection loss and noise in squeezing-based optical sensing'' contains sections on stabilizing the pump fluctuations (I), locking the phase (II), measurement of the SNL (III), the effect of the excess noise of the coherent beam (IV) and derivation of the phase sensitivity (V).

\nolinenumbers

\end{spacing}

\end{document}